\documentclass[twocolumn,showpacs,preprintnumbers,amsmath,amssymb]{revtex4}
\usepackage{graphicx}
\usepackage{dcolumn}
\usepackage{bm}

\usepackage{epsfig}
\newcommand{\beq}{\begin{equation}}
\newcommand{\eeq}{\vspace{0cm} \end{equation}}
\newcommand{\beqq}{\setlength\arraycolsep{2pt}\begin{eqnarray}}
\newcommand{\eeqq}{\vspace{0cm} \end{eqnarray}}
\newcommand{\e}{\textrm{\,e}}

\begin{document}

\title{Chemical Potential and the Nature of the Dark Energy: The case of phantom}

\author{J. A. S. Lima}\email{limajas@astro.iag.usp.br}

\author{S. H. Pereira} \email{spereira@astro.iag.usp.br}

\affiliation{Departamento de Astronomia, Universidade de S\~{a}o
Paulo \\ Rua do Mat\~ao, 1226 - 05508-900, S\~ao Paulo, SP, Brazil}


\bigskip
\begin{abstract}
The influence of a possible non zero
chemical potential $\mu$ on the nature of dark energy is investigated by assuming that the dark energy is a 
relativistic perfect simple fluid obeying the equation of state (EoS), $p=\omega \rho$ ($\omega <0, constant$).  
The entropy condition, $S \geq 0$,  
implies that the possible values of $\omega$ are heavily dependent on the magnitude, as well
as on the sign of  the chemical potential. For $\mu
>0$, the $\omega$-parameter must be greater than -1 (vacuum is
forbidden) while for $\mu < 0$ not only the vacuum but even a
phantomlike behavior ($\omega <-1$) is allowed. In any case, the
ratio between the chemical potential and temperature remains
constant, that is, $\mu/T=\mu_0/T_0$. Assuming that the dark energy
constituents have either a bosonic or fermionic nature, the general
form of the spectrum is also proposed. For bosons $\mu$ is always
negative and the extended Wien's law allows only a dark component
with $\omega < -1/2$ which includes vacuum and the phantomlike cases.
The same happens in the fermionic branch for $\mu <0$. However,
fermionic particles with $\mu >0$ are permmited only if $-1 < \omega <
-1/2$. The thermodynamics and statistical arguments constrain the
EoS parameter to be $\omega < -1/2$, a result surprisingly close to the
maximal value required to accelerate a FRW type universe dominated
by matter and dark energy ($\omega \lesssim -10/21$).
\end{abstract}

\maketitle

\section{Introduction}

The  current idea of an accelerating Universe is based on a large
convergence of independent observations, and its
explanation constitutes one of the greatest challenges for our
current understanding of fundamental physics \cite{SN,CMB}. In the context of a
Friedmann-Robertson-Walker (FRW) cosmology, dominated by
pressureless matter with density $\rho_m$ plus an extra component of
density $\rho$ and pressure $p$,  the scale factor evolution  is
governed by the equation 3${\ddot a}/a= -4\pi G(\rho_m + \rho +
3p)$. This means that a hypothetical component with large negative
pressure may drive the evolution of an expanding accelerating
Universe. This exotic component is usually termed dark energy or
quintessence, and it represents about 70\% of the total
content in the Universe. The origin and the nature of dark energy
is still a mystery, however, there is no doubt that its existence is
beyond the domain of the standard model of particle physics
\cite{review}.

Among a number of possibilities to describe the dark energy
component, the simplest and most theoretically appealing way is by
means of a cosmological constant $\Lambda$, which acts on the FRW
equations as an isotropic and homogeneous source with a constant
equation of state parameter $p/\rho=-1$. On the other hand, although
cosmological scenarios with a $\Lambda$ term might explain most of
the current astronomical observations, from the theoretical
viewpoint they are plagued with some fundamental problems thereby
stimulating the search for alternative dark energy models driven by
different candidates \cite{model,list1}. 

In the XCDM scenario, the dark energy component is
phenomenologically described by an equation of state $p = \omega\rho$. The
case $\omega = -1$ reduces to the cosmological constant.  
The imposition $\omega \geq -1$ is physically motivated by the 
classical fluid description
\cite{HElis82}. This hyphothesis introduces a strong bias 
in the $\omega$-parameter determination
from observational data. In order to circunvent such a
difficulty, superquintessence or phantom dark energy cosmologies
have been recently considered where such a condition is relaxed
\cite{faraoni02}. In contrast to the usual quintessence model, a
decoupled phantom component presents an anomalous evolutionary
behavior. For instance, the existence of future curvature
singularities, a growth of the energy density with the expansion, or
even the possibility of a rip-off of the structure of matter at all
scales are theoretically expected. Although possessing such strange
features, the phantom behavior is theoretically allowed by some
kinetically scalar field driven cosmology \cite{chi00}, as well as,
by brane world models \cite{Shani03}, and, perhaps, more important
to the present work, a PhantomCDM cosmology is not ruled out by the present
type Ia Supernovae and other observations \cite{ChouPadm,Alc04}. 

In a series of papers \cite{limamaia,LA04}, we have studied some
thermodynamics and statistical properties of dark energy with no
chemical potential ($\mu =0$). By using standard thermodynamics for
a relativistic simple fluid, we concluded that the case of phantom
energy is ruled out because the total entropy of a dark component with $\omega < -1$ is
negative. In addition, by combining thermodynamics and statistical
arguments  the EoS was restricted to the interval  $-1 \leq \omega <
-1/2$ and a fermionic nature to the dark energy particles was
favored.

Later on, thermodynamics arguments in favor of the phantom hypothesis
were put forward by Gonz\'alez-D\'{i}az and Sig\"uenza
\cite{gonzalez}. They claimed that the temperature of a phantomlike
fluid is always negative in order to keep its  entropy positive
definite (as statistically required). This new viewpoint was
justified by arguing that the scalar field representation of a
phantom field has a negative kinetic term ${\dot \phi}^{2}$ which
quantifies the translational kinetic energy of the associated fluid
system,  and, as such, its temperature (a measure of the average
kinetic energy) should be negative. Although temptative 
to some degree, both approaches have been considered in the literature 
(see, for instance, \cite{pavon,jorge} and Refs. therein).
More important to the present work, they share a
common property, namely, the chemical potential of the dark energy 
fluid was set to be zero from the very beginning.

In this article we reanalyze the thermodynamic and statistical
properties of the dark energy scenario by considering the existence
of a non-zero chemical potential. In order to clarify some
subtleties present in the earlier results, 
we rederive the physical quantities in 
the presence of $\mu$ by adopting the full
thermostatistic approach proposed in Refs. \cite{limamaia,LA04}.
This means that all thermodynamic and statistical properties follow
directly from the EoS plus the hypothesis that $\mu \neq 0$. In
particular, the temperatures  of the dark energy fluids
must be always positive definite.  This is an interesting
aspect of the present work since there are many scalar field
representations for dark energy fluids, however, the
thermodynamic laws are independent to what happens at a microscopic
level as long as the equation of state has been  defined.  As we
shall see, one of the main consequences of a negative chemical potential is 
that the phantom scenario is recovered without the need to
appeal to negative temperatures. In addition, a bosonic or fermionic
nature of the dark energy component now becomes possible.

The paper is planned as follows. In section 2 we discuss the
thermodynamic constraints when a chemical potential is introduced.
In  section 3 we consider a statistical analysis  by assuming that
the dark energy particles are massless and can have either a bosonic
or a fermionic nature. In the conclusion section, a joint analysis
involving the thermodynamic and statistical constraints on the
EoS $\omega$-parameter is presented

\section{Cosmology, dark energy and thermodynamics}

Let us now consider that the Universe is described by the
homogeneous and isotropic Friedmann-Robertson-Walker (FRW) geometry
($c=1$)
\begin{equation}
 ds^2 = dt^2 - a^{2}(t) \left(\frac{dr^2}{1 - \kappa r^{2}}
  + r^2 d \theta^2 +
 r^2 sin^{2}\theta d\phi^2\right),
\end{equation}
where $\kappa = 0,\pm 1$ is the curvature parameter and $a(t)$ is
the scale factor. In what follows we consider that the matter
content (or at least one of its noninteracting components is a fluid
described  by the EoS \beq p=\omega \rho\,,\label{eqstate} \eeq
where $p$ is the pressure, $\rho$ is the energy density and $\omega$
a constant parameter which may be positive (white energy) and
negative (dark energy).  The cases $\omega = 1/3$, $1$,  and $-1$
characterizes, respectively, the blackbody radiation, a stiff-fluid
and the vacuum state while $\omega < -1$ stands to a phantomlike
behavior.

Following standard lines, the equilibrium thermodynamic states of a
relativistic simple fluid are characterized by an energy momentum
tensor $T^{\alpha \beta}$, a particle current $N^{\alpha }$ and an
entropy current $S^{\alpha}$ which satisfy  the following relations
\begin{equation}
T^{\alpha \beta}=(\rho + p)u^{\alpha} u^{\beta} - pg^{\alpha \beta},
\quad T^{\alpha \beta};_{\beta}=0 , \label{eq:TAB}
\end{equation}
\begin{equation} \label{eq:NA}
N^{\alpha}=nu^{\alpha}, \quad  N^{\alpha};_{\alpha}=0 ,
\end{equation}
\begin{equation} \label{eq:SA}
S^{\alpha}=s u^{\alpha}, \quad  S^{\alpha};_{\alpha}=0 ,
\end{equation}
where ($;$) means covariant derivative, $n$ is the particle number
density, and $s$ is the entropy density. In the FRW background, the
above conservation laws can be rewritten as (a dot means comoving
time derivative)

 \begin{equation}
 \dot{\rho} + 3 (1 + \omega)\rho \frac{\dot a}{a}=0,\,\,
 \dot{n} + 3n\frac{\dot a}{a}=0,\,\, \dot{s} + 3s\frac{\dot a}{a}=0,
\end{equation}
whose solutions can be written as:

\begin{equation}
\rho=\rho_0 \left(\frac{a_0}{a} \right)^{3(1 + \omega)}, \,\, n=n_0
\left(\frac{a_0}{a}\right)^{3}, \,\, s=s_0
\left(\frac{a_0}{a}\right)^{3},\label{eqSol}
\end{equation}
where the positive constants $\rho_0$, $n_0$, $s_0$ and $a_0$ are
the present day values of the corresponding quantities (hereafter
present day quantities will be labeled by the index ``0"). On the other
hand, the quantities $p$, $\rho$, $n$ and $s$ are related to the
temperature $T$ by the Gibbs law

\begin{equation} \label{eq:GIBBS}
nTd(\frac{s}{n})= d\rho - {\rho + p \over n}dn,
\end{equation}
and from Gibbs-Duhem relation \cite{callen} there are only two
independent thermodynamic variables, say, $n$ and $T$. Now, by
assuming that $\rho=\rho(T,n)$ and $p=p(T,n)$ and combining  the
thermodynamic identity \cite{weinb}

\begin{equation}
T \biggl({\partial p \over \partial T}\biggr)_{n}=\rho + p - n
\biggl({\partial \rho \over \partial n}\biggr)_{T},
\end{equation}
with the conservation laws as given by (6), one may show that the
temperature satisfies
\begin{equation} \label{eq:EVOLT}
{\dot T \over T} = \biggl({\partial p \over \partial
\rho}\biggr)_{n} {\dot n \over n} = -3\omega \frac{\dot a}{a}.
\end{equation}
Therefore, assuming that $\omega \neq 0$ we obtain

\begin{equation} \label{eq:TV}
n = n_0 \left(\frac{T}{T_0} \right)^{1 \over  \omega}  \quad
\Leftrightarrow \quad T=T_0 \left(\frac{a}{a_0} \right)^{-3\omega}.
\end{equation}
The temperatures appearing in the above expressions are positive
regardless of the value of $\omega$.  In particular, in the
radiation case ($\omega = 1/3$), one finds $aT=a_0T_0$ as should be
expected. As compared to this case, the unique difference is that
the dark energy fluid (even in the phantom regime) becomes hotter in
the course of the cosmological adiabatic expansion since its
equation-of-state parameter is a negative quantity. A physical
explanation for this behavior is that thermodynamic work is being
done on the system \cite{limamaia,LA04}.

It should  be stressed that the derivation of the temperature
evolution law presented here is fully independent of the entropy
function, as well as, of the chemical potential $\mu$. The above
expressions also imply that for  any comoving volume of the fluid,
the product $T^{1 \over  \omega} V$  remains constant in the course
of expansion and must also characterize the equilibrium states
(adiabatic expansion) regardless of the value of $\mu$. Further, by
inserting the temperature law into the energy conservation law
(\ref{eqSol}), one obtains the energy density as function of the
temperature

\begin{equation}
\rho= \rho_0 \left(\frac{T}{T_0} \right)^{\frac{1 +
\omega}{\omega}}.
\end{equation}

Now, in order to determine the chemical potential and its influence
on the thermodynamics of dark energy,  we consider the Euler
relation \cite{callen}
\beq Ts={p+\rho}-{\mu n},\,\label{entropyd}
\eeq
where $\mu$ in general can also be a function of $T$ and $n$
\cite{FUG,degroot}. By combining the above expression with equations
(\ref{eqstate}), (\ref{eqSol})  and (\ref{eq:TV}) we obtain:

\beq \mu=\mu_0\left(\frac{a}{a_0} \right)^{-3\omega} =
\mu_0\left(\frac{T}{T_0} \right), \eeq where \beq\label{mu0}
\mu_0={1\over n_0}[(1+\omega)\rho_0-T_0 s_0].
\eeq

\begin{figure}[htb]
\begin{center}
\epsfig{file=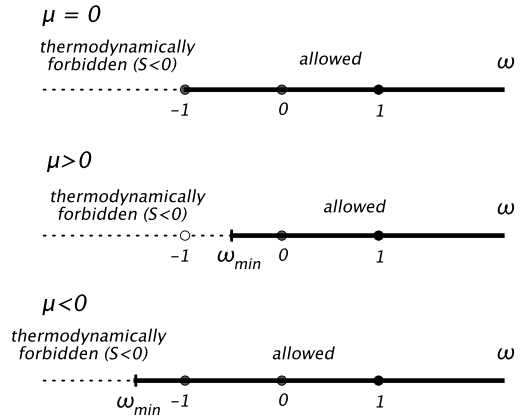, scale=0.6}\\
\end{center}
\caption{The allowed intervals of $\omega$ values (heavy lines) and
forbidden (dashed lines) for null, positive and negative chemical
potentials. Note that a large portion of the dark branch $\omega <
0$ is always  thermodynamically permitted. However,  for $\mu \geq
0$, the phantomlike behavior ($\omega<-1$) is thermodynamically
forbidden. }\label{figt}
\end{figure}

This straightforward thermodynamic result has some interesting
consequences.  In principle, the chemical potential may be either
positive or negative, and it also  depends on the values of the EoS
$\omega$-parameter. In particular,  $\mu$ is always negative ($\mu_0
< 0$) in the case of phantom  energy, and becomes even more
negative in the course of time ($T$ grows with the scale factor
during the cosmic evolution). It is also known that $\mu$ is zero in
the case of photons ($\omega=1/3$) because they are their own
antiparticles \cite{landau}.  In this case, (\ref{mu0}) yields
correctly that  $3s_0 T_0=4\rho_0$ as should be expected. In
general, if $\mu=0$, necessarily  the  relation $s_0 T_0= (1 +
\omega)\rho_0$ must be obeyed, which is just the present day
expression of $sT=(1 + \omega)\rho$ as required by (\ref{entropyd}).

At this point, the fundamental question is: How the chemical
potential modifies the entropy constraints \cite{limamaia,LA04}
derived in the previous papers?

\begin{figure*}
\centerline{\psfig{figure=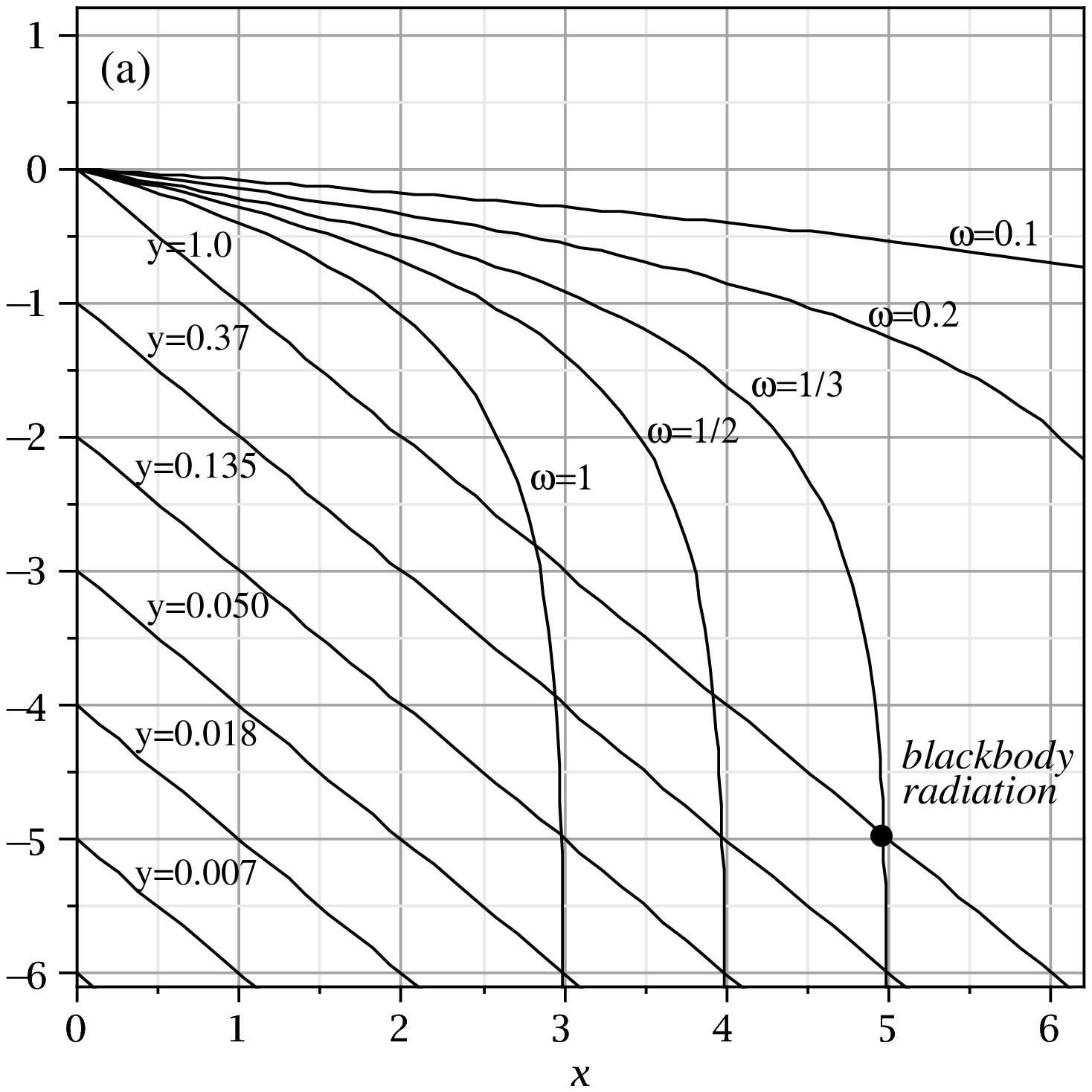,width=3.1truein,height=2.5truein}
\psfig{figure=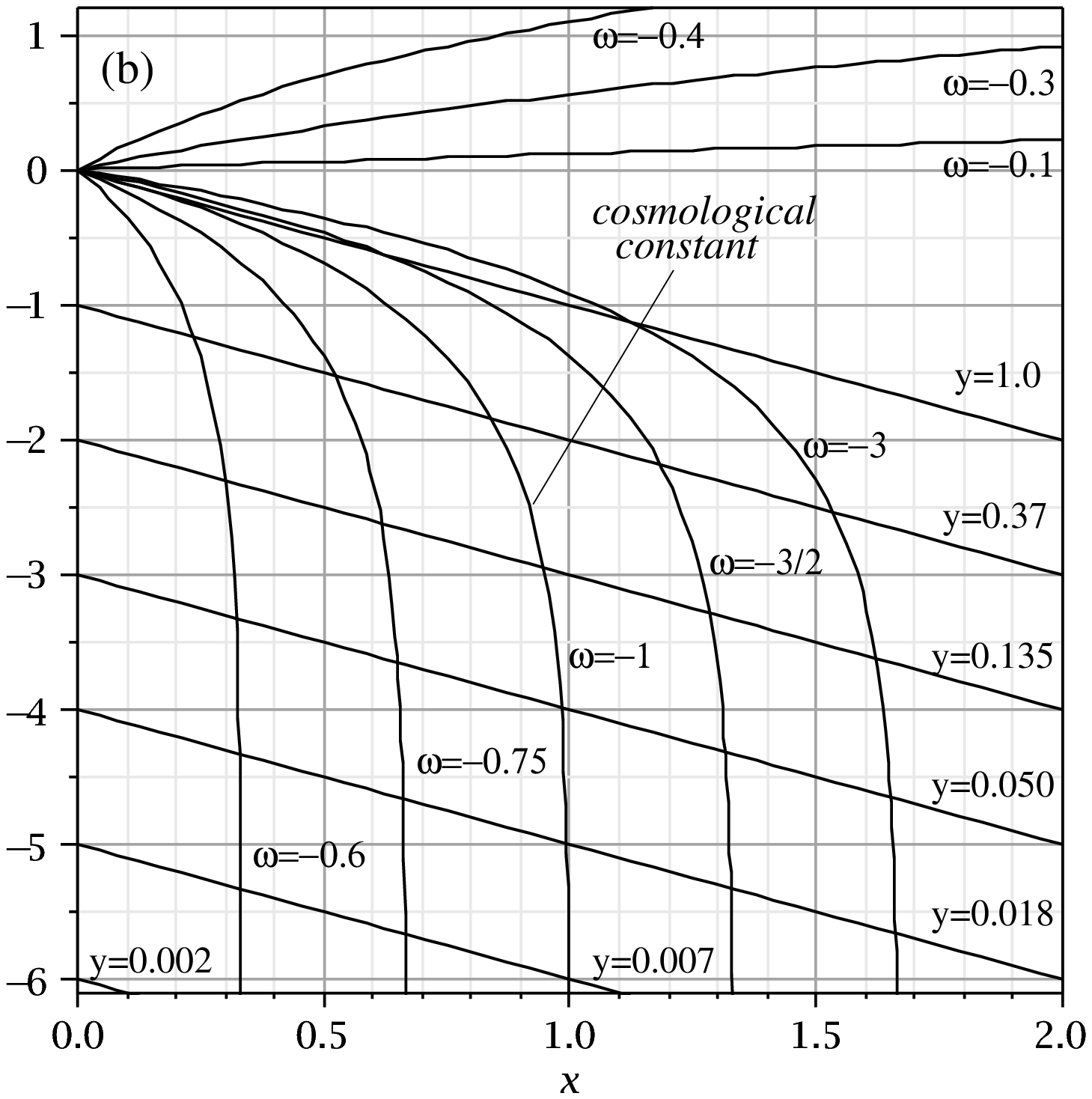,width=3.1truein,height=2.5truein} \hskip
0.1in} \caption {Solutions for the bosonic case ($\mu \leq 0$).
{\bf{a)}} For given values of the pair (x,y), the straight lines and
curves represent the l.h.s. side and the r.h.s. of
(\ref{conditionB}), respectively. The intersection points between
the curves and the straight lines represent the desired physical
solutions for $\omega>0$. The standard radiation solution ($\mu=0$
and $\omega=1/3$) is indicated by a black point. {\bf{b)}} Curves
representing the right hand side of (\ref{conditionB}) for different
$\omega$ values in the region $\omega<-1/2$ (lower half plane) and
$-1/2<\omega \leq 0$ (upper half plane). The cosmological constant
case ($\omega = -1$) is indicated. For both diagrams the fugacity
$y=\exp(\mu/k_B T) \leq 1$.}
\end{figure*}

In order to show that we compute explicitly the entropy of dark
energy for a comoving volume $V$. As remarked before, the entropy
function should scale as $S \propto T^{1 \over \omega}V$. Actually,

\beq S(T,V)\equiv sV = \left[\frac{(1+\omega)\rho_0-\mu_0
n_0}{T_0}\right] \bigg({T\over
T_0}\bigg)^{1/\omega}V=s_0V_0,\label{entropy} 
\eeq 
which remains
constant as expected (see discussion below Eq.(\ref{eq:TV})).
However, in order to keep the entropy $S\geq 0$ (as statistically
required), the following constraint must be satisfied: \beq \omega
\geq \omega_{min}=-1 + {\mu_0 n_0 \over \rho_0}\,,\label{accord}
\eeq which introduces a minimal value to the $\omega$-parameter,
below which the entropy becomes negative. This is a remarkable
expression and its consequences are apparent. For instance, consider
that $\mu_0=0$ (no chemical potential). In this case,  the smallest
value of the $\omega-parameter$ is $\omega_{min}=-1$ and the
previous analysis by Lima and Alcaniz \cite{LA04} is fully
recovered, that is, the phantom domain ($\omega < -1$) is
thermodynamically forbidden. However, for a negative chemical
potential, the phantomlike regime becomes thermodynamically allowed
thereby recovering the hypothesis of phantom energy without
appealing to negative temperature as proposed in the literature
\cite{gonzalez}.  Note also that for a positive chemical potential
not even a cosmological constant ($\omega = -1$) is possible. In
figure 1, we summarize the basic thermodynamic results.

\section{Dark energy with a chemical potential: Statistical behavior}

Another interesting feature of a dark energy component with  a non
zero chemical potential is related to its spectral distribution. The
generalized Wien-type spectrum for dark energy with $\mu=0$  has
already been discussed in the literature \cite{limamaia,LA04}. A
different approach for the phantomlike behavior involving the
modulus of the temperature has also  been proposed  \cite{gonzalez}.
In the present case, since the  temperature is positive,  we simply
add the chemical potential $\mu$ to the spectrum previously derived
\cite{limamaia,LA04}. More precisely, we postulate the following
spectral distribution: \beq \rho(T,\nu)={\alpha\nu^{1/\omega}\over
\e^{(h\nu-\mu)/k_B T}+\epsilon}\,,\label{spectr} \eeq where
$\epsilon=+1$ stands for the Fermi-Dirac distribution and
$\epsilon=-1$ to the Bose-Einstein one, and $\alpha$ is  a positive
and $\omega$-dependent constant. Here it is important to note that
for bosons the chemical potential is always negative or null, while
for fermions it may be positive or negative \cite{landau}.

A direct consequence of (\ref{spectr}) is related to the
displacement Wien's law. The wavelength for which the distribution
attains its maximum value is determined by the condition

\begin{equation}\label{roots}
\lambda_m T = {hc \over k_{B}x'(\omega)} = {1.438 \over x'(\omega)},
\end{equation}
where $x'(\omega)$ is the root of the  transcendent equation

\beq y\e^{-x}=-\epsilon + \left[\epsilon \omega
\over{1+2\omega}\right] x\,,\label{condition}
\eeq
where $x=hc/k_B
\lambda T$ and $y=\exp(\mu/k_B T)\equiv \exp(\mu_0/k_B T_0)$ is a
constant fugacity. When $\mu_0=0$ the above expressions reduces to
the one obtained in \cite{LA04}. The solution of the above algebraic
equation can be derived both numerically and graphically. We are
only interested in solutions with positive $x$, because the
temperature is always positive. Due to the presence of the chemical
potential, the analysis of the above condition will be done
separately for bosons and fermions.

\subsection{Bosons}

Let us now analyze the bosonic case ($\epsilon=-1$). It proves
convenient  to rewrite condition (\ref{condition}) in the following
form:

\beq \ln y -x =\ln \bigg(1- \frac{\omega
x}{1+2\omega}\bigg)\,.\label{conditionB}
\eeq

For each value of $\mu$, the left hand side (l.h.s.) of the above
expression is a collection of straight lines with  slope equal to
$-1$. Since $\mu$ is always negative or null for bosons, it follows
that $0<y\leq 1$ so that $-\infty < \ln y \leq 0$. This means that
the l.h.s. of (\ref{conditionB}) is a collection of parallel
straight lines on the lower half plane. The first straight line is
the trivial solution with zero chemical potential.

\begin{figure}[htb]
\begin{center}
\epsfig{file=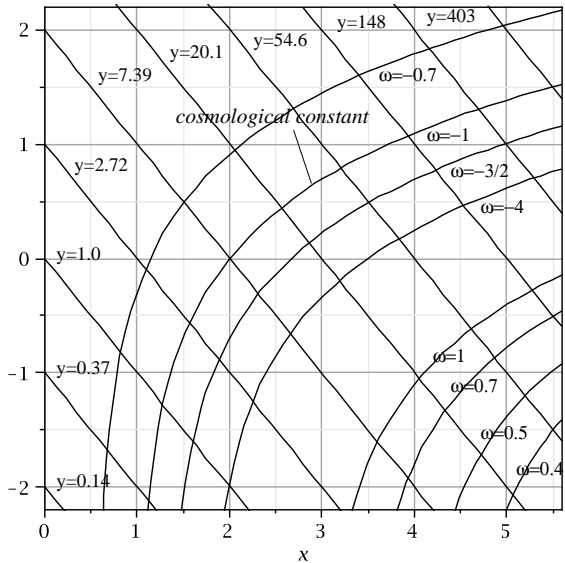, scale=0.5}
\end{center}
\caption{Curves representing the right hand side of
(\ref{conditionF}), for different $\omega$ values in the region
$\omega<-1/2$ and $\omega> 0$.}\label{fig3}
\end{figure}

Note also that
the right hand side (r.h.s.) of (\ref{conditionB}) involves the
singularity for $\omega = -1/2$, and, as such, must be separately
analyzed. Apart from this point, we have 3 different intervals,
namely: $\omega > 0$, $\omega < -1/2$ and $-1/2<\omega \leq 0$.

In Figures 2a and 2b we display the main results. When $\omega>0$ we
have a collection of curves represented in Fig. 2a. All of them
cross the straight lines in some point $x>0$ thereby indicating the
solutions of the algebraic equation (\ref{conditionB}). The standard
radiation solution ($\mu=0$ and $\omega=1/3$) is indicated, however,
it should also be remarked the theoretical possibility of a
radiation solution with $\mu \neq 0$. For $\omega < -1/2$ we have
the collection of curves represented on the lower half plane of
Figure 2b, superposed to the straight lines. This analysis show that
all these curves cross the straight lines in some positive $x$
value, indicating a solution to the algebraic expression
(\ref{conditionB}). Finally, on the interval $-1/2<\omega \leq 0$ we
have the set of curves represented on the upper half plane of Figure
2b thereby showing the absence of physical solutions.

In summ, a simple graphic analysis shows that there are two
intervals of $\omega$ for which the condition (\ref{conditionB}) has
a solution, namely, $\omega > 0$ and $\omega < -1/2$, while the
interval  $-1/2 < \omega \leq 0$ is statistically forbidden.
Therefore, unlike the previous analyzes with $\mu=0$
\cite{limamaia,LA04},  the EoS $\omega < -1/2$ for bosons now
becomes possible when the chemical potential is negative. This
includes the phantom dark energy as a physical possibility.

\subsection{Fermions}

The analysis of the fermionic case ($\epsilon=+1$) is similar to the
bosonic one, but now the chemical potential can be either positive
or negative. The condition (\ref{conditionB}) for the fermionic case
reads \beq \ln y -x =\ln \bigg(-1 + \frac{\omega
x}{1+2\omega}\bigg).\label{conditionF} \eeq

\begin{figure}\label{fig4}
\begin{center}
\epsfig{file=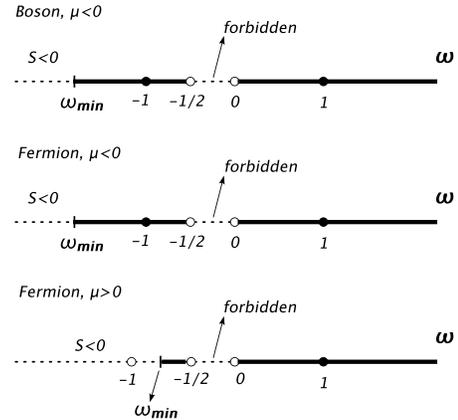, scale=0.55}
\end{center}
\caption{Thermodynamic and statistical constraints. The allowed
(heavy lines) and forbidden (dashed lines) values of $\omega$ for
the bosonic and fermionic cases with $\mu<0$ and $\mu>0$. The
phantom branch $\omega < -1$ is excluded for the fermionic case with
$\omega > 0$. Note that the dark branch $-1<\omega<-1/2$ for bosons
is now possible.}
\end{figure}

The analysis on the l.h.s. of (\ref{conditionF}) is similar to the
bosonic case, the unique difference is that $\mu_0$ can be positive.
In this case, we have $y>1$ and $\ln y > 0$. As in the previous
case, the discussion on the r.h.s. of (\ref{conditionF}) depends on
the $\omega$ values. For the cases $\omega < -1/2$ and $\omega>0$ we
have the curves represented in Figure 3. We see that all curves
crossing the straight lines for some positive value of $x$ yield a
valid solution for (\ref{conditionF}). Note also that on the
interval $-1/2<\omega<0$, all the curves are in the negative
$x$-axis (negative temperatures), and, therefore, none of them cross
the straight lines (two reasons for the the interval be a forbidden
region).

\section{Concluding Remarks}

In this paper we have investigated the thermodynamic and statistical
properties of a dark energy fluid with equation of state, $p=\omega
\rho$, by assuming that its chemical potential is different from
zero.

In Figure 4, we summarize the main results of the present analysis
by combining both approaches.  As discussed in section 2, the
regions with $S < 0$ are  thermodynamically forbidden. Note also
that  many dark energy fluids satisfy the combined constraints
regardless of the $\mu$ sign, that is, a large interval of  negative
$\omega$ values is allowed by thermodynamic and statistical
considerations.  However, a phantom like behavior ($\omega < -1$) is
permitted only for $\mu < 0$, and the corresponding massless
particles can have either a bosonic or fermionic nature.

It was also proved (see also Fig. 4) that the EoS parameter of a
dark energy fluid obeying a generalized Wien's law always satisfies
the constraint $\omega<-0.5$ (a thermostatistics limit).  This upper
limit is surprisingly close to the maximal value of the EoS
$\omega$-parameter  necessary to accelerate the present universe.
Actually, in order to accelerate  a FRW universe dominated by matter
and dark energy, the EoS parameter must satisfies the inequality,
$\omega <  -(1+ \Omega_m/\Omega_x)/3$. Therefore, for $\Omega_m\cong0.3$
and $\Omega_x\cong0.7$, as indicated by the present observations
\cite{SN,CMB}, one finds the dynamic constraint $\omega \lesssim - 10/21$.

Finally, it should be stressed that for $\mu=0$ one finds  $\omega_{min}=-1$
(see Eq. (\ref{accord})) in accordance to the results previously
derived by Lima and Alcaniz \cite{LA04}. The present analysis with
$\mu \neq 0$  also opens  the possibility for an EoS parameter $\omega <
-1$, thereby recovering the idea of a phantom dark energy without to
appeal to negative temperatures. Perhaps more interesting, unlike
the results  for $\mu=0$  which favored only a fermionic nature to
the dark energy fluid (phantom and nonphantom), it was demonstrated
that a bosonic kind of dark energy becomes possible from a
thermostatistics viewpoint.

\begin{acknowledgments}
The authors would like to thank V. C. Busti, J. V.
Cunha, J. F. Jesus, A. C. Guimaraes, R. Holanda and R. C. Santos for helpful
discussions. JASL is partially supported by CNPq and FAPESP No.
04/13668-0 and SHP is supported by CNPq No. 150920/2007-5 (Brazilian
Research Agencies).
\end{acknowledgments}


\begin{thebibliography}{99}
\bibitem{FUG} In the case of a relativistic ideal gas of massive particles, for instance,
the so-called  fugacity reads, $exp(\mu/k_B T) = n(2\pi\bar
h)^{3}/4\pi m^{2}ck_B K_2(mc^{2}/k_B T)$, where $m$ is the particles
mass and $K_2$ is the modified  Bessel of second kind.
\end{thebibliography}
\end{document}